%% file: Nonconvex.tex
\documentclass[]{article}  
\usepackage{url}
\usepackage{graphicx}
\usepackage{amsfonts}
\usepackage{amssymb}
\usepackage{latexsym}

\input{mac}

\def\g{{\gamma}}
\def\t{{\theta}}
\def\d{{\delta}}
\def\e{{\varepsilon}}
\def\D{{\Delta}}

\newcommand{\notyet}[1]{}

\begin{document}

\title{Partitioning Regular Polygons into Circular Pieces II:\\
Nonconvex Partitions}

\author{%
Mirela Damian%
   \thanks{Department of Computer Science, Villanova University, Villanova,
PA 19085, USA.
   \protect\url{mirela.damian@villanova.edu}.}
\and
  Joseph O'Rourke%
    \thanks{Department of Computer Science, Smith College, Northampton, MA
      01063, USA.
      \protect\url{orourke@cs.smith.edu}.
       Supported by NSF Distinguished Teaching Scholars award
       DUE-0123154.}
}

\date{}
\maketitle
\begin{abstract}
We explore optimal circular nonconvex partitions of regular $k$-gons. 
The \emph{circularity} of a polygon is measured by its 
\emph{aspect ratio}: the ratio of the diameters of the smallest 
circumscribing circle to the largest inscribed disk. 
An optimal circular partition minimizes the maximum ratio over all
pieces in the partition.
We show that the equilateral triangle has an optimal $4$-piece 
nonconvex partition, the square an optimal $13$-piece nonconvex 
partition, and the pentagon has an optimal nonconvex partition with 
more than $20$ thousand pieces.
For hexagons and beyond, we provide a general algorithm that
approaches optimality, but does not achieve it.
\end{abstract}

\section{Introduction}
In~\cite{mo-cp-03} we explored partitioning regular $k$-gons into
``circular'' convex pieces.
Circularity of a polygon is measured by the 
\emph{aspect ratio}: the ratio of
the diameters of the smallest circumscribing circle to the largest
inscribed disk.  We seek partitions with aspect ratio close to $1$,
ideally the optimal ratio.
Although we start with regular polygons, most of the machinery developed
extends to arbitrary polygons.

For convex pieces, we showed in~\cite{mo-cp-03}
that optimality can be achieved for an equilateral triangle only
by an infinite partition, and that for all $k \ge 5$, the $1$-piece partition is
optimal. We left the difficult case of a square unsettled, narrowing
the optimal ratio to a small range.
Here we turn our attention to partitions that permit the pieces to
be nonconvex.
The results are cleanest if we do not demand that the pieces be polygonal,
but rather permit curved sides to the pieces.
The results change dramatically compared to the convex case.
The equilateral triangle has an optimal $4$-piece partition,
the square an optimal $13$-piece partition,
the pentagan an optimal partition with more than $20$ thousand pieces.
For hexagons and beyond, we provide a general algorithm that approaches
optimality, but does not achieve it.

\subsection{Notation}
A \emph{nonconvex partition} of a polygon $P$ is a collection 
of sets $S_i$ satisfying
\begin{enumerate}
\item Each $S_i \subseteq P$.
\item $\cup_i S_i = P$.
\item The sets have pairwise disjoint interiors.
\end{enumerate}
These conditions alone are too broad for our purposes, as there are no
constraints placed on the pieces. 
It is reasonable to demand that each set be connected, but even
this is too broad.  The most natural constraint for our purposes
is to require the interior of each piece to be connected:
\begin{enumerate}
\setcounter{enumi}{3}
\item The interior of each $S_i$ is connected.
\end{enumerate}
The aspect ratio of a piece is the ratio of the radius of the smallest
circumcircle to the radius of the largest inscribed disk. 
Aspect ratios will be denoted by symbol $\g$, modified by subscripts and
superscripts as appropriate: $\g_1(P)$ is the one-piece $\g$; 
$\g(P)$ is the maximum of all the $\g_1(S_i)$ for all pieces $S_i$ in
a partition of $P$; 
$\g^*(P)$ is the minimum $\g(P)$ over all nonconvex partitions of $P$.
Our goal is to find $\g^*(P)$ for the regular $k$-gons $P$.
Both the partition and the argument ``$(P)$'' will often be dropped
when clear from the context.

Throughout we consider all disks to be closed sets, including
the points on their bounding circle.
Disks will be denoted either by symbols $D^i$, $i=1,2,3,\ldots,n$;
the subscript $0$ will indicate the disk bound by an inscribed/in-disk,
and $1$ will indicate the circumscribed/out-circle.

\subsection{Table of Results}
\seclab{Results}
Our results are summarized in Table~\tabref{Results}.

\begin{table}[htbp]
\small
\begin{center}
\begin{tabular}{| l | c | c | c | c|}
	\hline
\mbox{}
        & \mbox{}
        & \mbox{}
        & \multicolumn{2}{c |}{~~~~~~nonconvex, ~~nonpolygonal~~~~~~} 
        \\ \cline{4-5}
\emph{Polygon}
	& $\g_1$
	& $\g_\t$
        & $\g^*$
        & $k^*$
       	\\ \hline \hline
Equilateral Triangle
	& $2.00000$
	& $1.50000$
        & $\g_\t$
        & $4$
       	\\ \hline
Square
	& $1.41421$
	& $1.20711$
        & $\g_\t$
        & $13$
       	\\ \hline
Regular Pentagon
	& $1.23607$
	& $1.11803$
        & $\g_\t$
        & $\le 20476$
       	\\ \hline
Regular Hexagon
	& $1.1547$
	& $1.07735$
	& $1.10418$
        & finite
       	\\ \hline
Regular Heptagon
	& $1.10992$
	& $1.05496$
	& 1.08382
        & finite
       	\\ \hline
Regular Octagon
	& $1.08239$
	& $1.0412$
	& $\g^8_1=1.08239$
        & finite
       	\\ \hline \hline
Regular $k$-gon
	& $~~~~1/\cos ( \pi/k )~~~~$
	& $~~~~\frac{1 + \csc( \t/2 )}{2}~~~~$
        & $\le \g^8_1=1.08239$
       & finite
        \\ \hline \hline
\end{tabular}
\caption{Table of Results on Regular Polygons.
$\g_1$: one-piece partition; 
$\g^k_1$: one piece ratio $\g_1$ for regular $k$-gon;
$\g_\t$: single-angle lower bound; $\t$: angle at corner;
$\g^*$: optimal partition; $k^*$: number of pieces in optimal partition.
}
\medskip
\noindent
\end{center}
\normalsize
\tablab{Results}
\end{table}
\noindent
Here we use $\g_\t$ to denote the ``one-angle lower bound'', a lower 
bound derived from one angle $\t$ of the polygon, ignoring all else.  
This presents a trivial lower bound on the aspect ratio of any 
partition.

\section{Preliminary lemmas}
We recall two simple lemmas used in Table~\tabref{Results},
proved in~\cite{mo-cp-03}.

\begin{lemma} {\em \bf (Regular Polygon).}
The aspect ratio $\g_1$ of a regular $k$-gon is 
 \[ \g_{1} =  \frac{1}{\cos(\pi/k)} \]
\lemlab{Reg.Poly}
\end{lemma}

\begin{lemma} {\em \bf (One-Angle Lower Bound).}
If a polygon $P$ contains a convex vertex of internal angle $\t$, 
then the aspect ratio of a partition of $P$ is no smaller than
$\g_{\t}$, with
 \[ \g_{\t} =  \frac{1 + \csc(\t/2)}{2} \]
\lemlab{one.angle.lower.bound}
\end{lemma}

\section{Equilateral Triangle}
An equilateral triangle has $\g_1 = 2$.
The lower bound provided by Lemma~\lemref{one.angle.lower.bound}
is $\g_\t = 1.5$ (see Table~\tabref{Results}). 
Figure~\figref{partition.triangle} shows a partition with 
$4$ pieces that achieves $\g_\t$, and is therefore optimal.
This partition has three convex corner pieces 
and one nonconvex central piece.

\begin{figure}[htbp]
\centering
\includegraphics[width=0.3\linewidth]{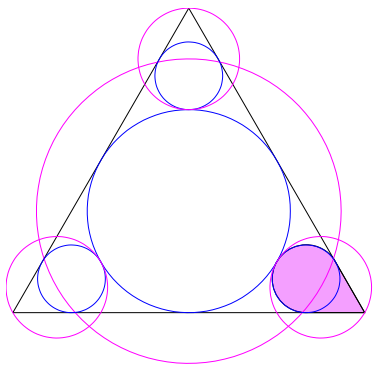} 
\caption{Optimal partition of an equilateral triangle (4 pieces).
Inscribed and circumscribed circles are shown.}
\figlab{partition.triangle}
\end{figure}

\section{Square}
A square has $\g_1 = \sqrt{2} \approx 1.41421$.
The lower bound provided by Lemma~\lemref{one.angle.lower.bound}
is $\g_\t = (1+\sqrt{2})/2 \approx 1.20711$ 
(see Table~\tabref{Results}). 
Figure~\figref{partition.square}a shows a partition with 
$13$ pieces that achieves $\g_\t$, and is therefore optimal.
\begin{figure}[htbp]
\centering
\begin{tabular}{c@{\hspace{0.7in}}c}
\includegraphics[width=0.4\linewidth]{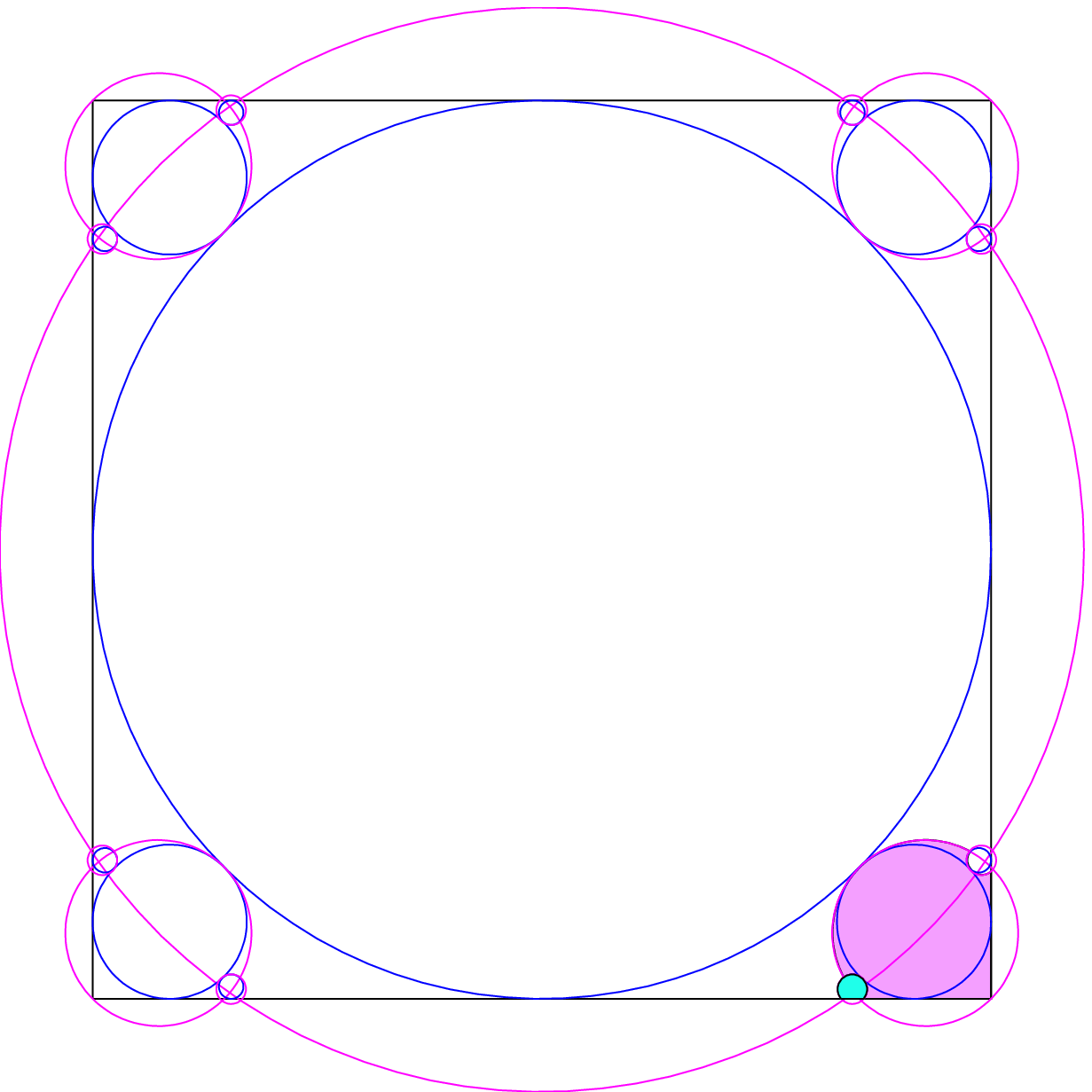} &
\includegraphics[width=0.38\linewidth]{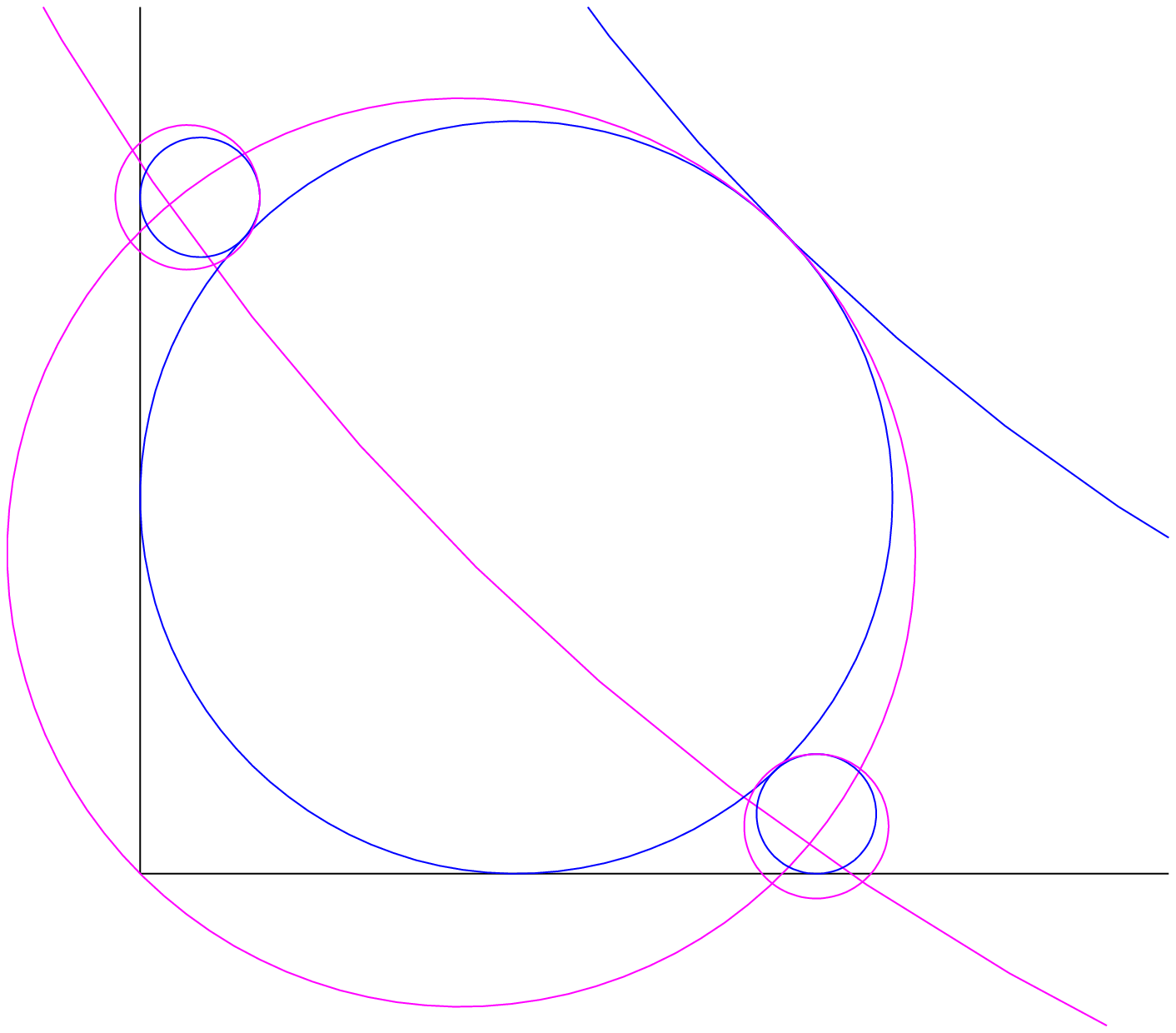} \\
(a) & (b) 
\end{tabular}
\caption{(a) Optimal partition of a square (13 pieces)
(b) Magnified view of one square corner.}
\figlab{partition.square}
\end{figure}
The partition contains one large central nonconvex piece, four
convex corner pieces and one nonconvex piece to each side of each
corner piece, for a total of 13 pieces.

As $k$ increases, $\g_\theta$ decreases and it becomes increasingly
difficult to partition a $k$-gon into pieces with optimal ratio.
As hinted in the square partition, it becomes essential to be able
to cover small gaps along the interior of edges.
Even for the pentagon, a less ad hoc procedure is needed.
In the next section, we
devise a general algorithm that covers a subsegment of an edge
with pieces with ratio close to optimal. 
This will permit us to make
progress for $k > 4$.

\section{Covering an edge segment}
Let $S$ be an edge segment tangent to two disks $D_0^0$ and 
$D_0^1$ at its endpoints.
A {\em covering} of $S$ is two collection of disks, 
$D_0^i$ and $D_1^i$, $i=0,1,2,3,\ldots $ with 
four properties:
\begin{enumerate}
\item Each disk $D_0^i$ is tangent to $S$
\item The disks $D_0^i$ have pairwise disjoint interiors:
$\int(D_0^i) \cap \int(D_0^j) = \emptyset$ for $i \neq j$.
\item The disks $D_1^i$ collectively cover $S$:
$\cup_i D_1^i \supseteq S$.
\item Each $D_0^i$ is inside the corresponding $D_1^i$:
$D_0^i \subseteq D_1^i$ for all $i$.
\end{enumerate}
For a given edge segment $S$, our goal is to find a covering
of $S$ of optimal ratio. In 
the following we present an algorithm that finds
a covering of $S$ of ratio close to the optimal. 

\subsection{Algorithm (Edge Cover)}
\seclab{triangular.gap}
The algorithm presented in this section takes as input:
\begin{itemize}
\item [(a)] An edge segment $S = [a_0, a_1]$ 
\item [(b)] Disks $D_0^0$ and $D_0^1$ tangent to each other and to $S$ at points 
$a_0$ and $a_1$, respectively
\item [(c)] Corresponding outcircles $D_1^0 \supset D_0^0$ and $D_1^1 \supset D_0^1$ 
\item [(d)] The desired ratio factor $\g > 1$
\end{itemize}
and seeks to extend the sets $\{D_0^0, D_0^1\}$ and 
$\{D_1^0, D_1^1\}$ to a {\em covering} of $S$ of ratio $\g$, 
if one exists. 

For $i \in \{0, 1\}$, let $a_i$ be the point where $D_0^{i}$ 
touches $S$ and $b_i$ the point where $D_1^{i}$ intersects $S$. 
We start by growing the largest possible indisk $D_0^2$ 
that touches the uncovered segment piece at midpoint 
$a_2 = (b_0 + b_1)/2$. Clearly, 
$D_0^2$ can only grow until it touches either of the 
two adjacent indisks, $D^0_0$ or $D^1_0$.
We will show later that $D_0^2$ hits the smaller of
$D_0^0$ and $D_0^1$ first 
(see Figure~\figref{gap.algorithm}a). 
Next we inflate $D_0^2$ by $\g$ to obtain $D_1^{2}$ and 
displace $D_1^{2}$ vertically downwards until its topmost 
point touches the topmost point of $D_0^2$, so as to capture as 
much of the uncovered edge segment as possible.

\begin{figure}[htbp]
\centering
\begin{tabular}{c@{\hspace{0.8in}}c}
\includegraphics[width=0.45\linewidth]{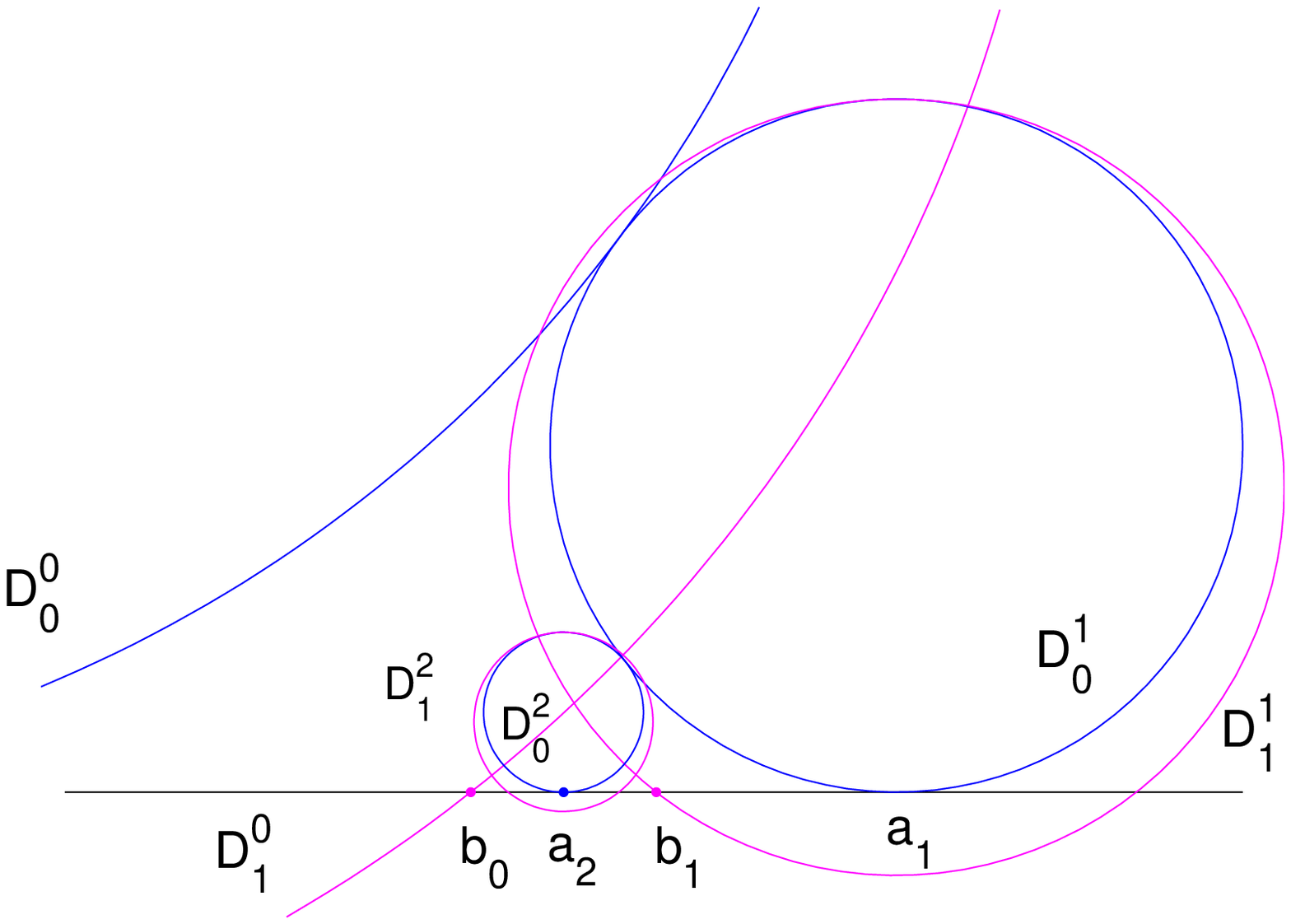} &
\includegraphics[width=0.39\linewidth]{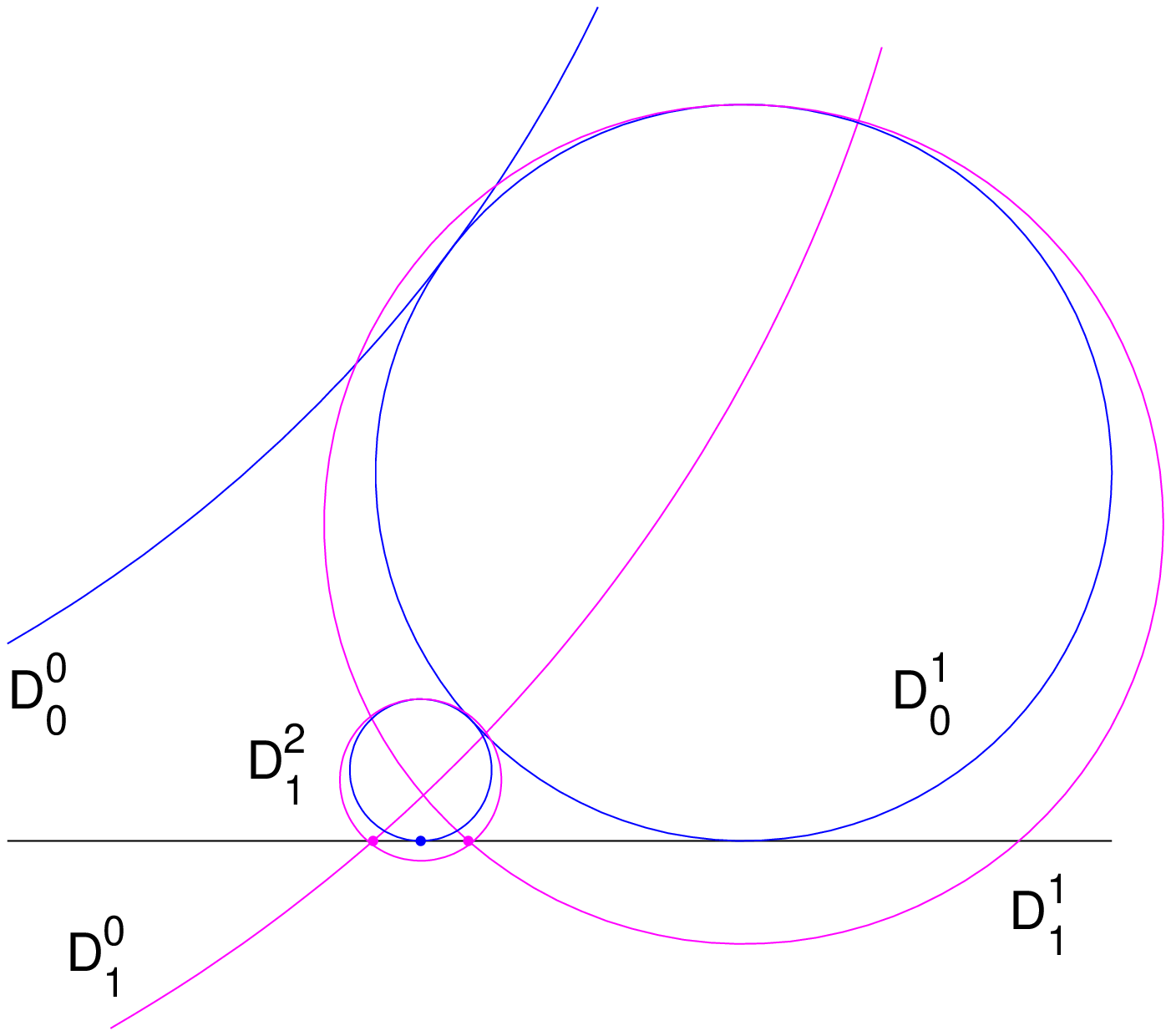} \\
(a) & (b)  
\end{tabular}
\caption{Algorithm (a) Iterative step: $D_0^2$ centered on the 
midpoint $a_2$ of $[b_0, b_1]$ (b) Termination: $D_1^2$ covers 
the gap.}
\figlab{gap.algorithm}
\end{figure}

If $D_1^2$ covers the entire triangular gap 
(as in Figure~\figref{gap.algorithm}b),  we are finished.
Otherwise, recurse on the at most two new edge segments 
created: $[a_0, a_2]$ and $[a_2, a_1]$.
Note that the uncovered gaps of these two edge segments are 
identical and therefore their coverings will be identical.  

\subsection{Analysis}
\seclab{analysis}
Without loss of generality, we assume that $D_0^0$ is at least 
as large as $D_0^1$. 
For analysis convenience, consider a coordinate system with the 
origin where $D_1^0$ intersects the horizontal edge, as 
in Figure~\figref{gap.proof}. 
At a certain stage of the algorithm, all uncovered gaps in the 
original edge segment are symmetric and will be covered in 
the same way. In our analysis, we focus on the uncovered gap 
adjacent to the origin; henceforth, the term gap will refer to the 
leftmost uncovered gap of the edge segment, with leftmost
understood.

Let $D^n_0$ be the indisk on the right side of the gap in 
iteration step $n$; $D^0_0$ always remains to 
the left side of the gap. 
Refer to Figure~\figref{gap.proof}.
For any $n$, let $r_n$ denote the radius of $D^n_0$ and 
let $a_n$ be the point where $D^n_0$ touches the $x$-axis.
We define a useful quantity $\d_n$ to represent the distance 
from $a_n$ to where $D^n_1$ intersects the $x$-axis: 
$\d_n = r_n \sqrt{\g^2 - (2-\g)^2}$, or equivalently 
\begin{equation}
\d_n = 2 r_n \sqrt{\g-1} 
\eqlab{delta}
\end{equation}
In iteration step $n+1$, the algorithm grows the indisk $D^{n+1}_0$ 
tangent to the uncovered gap $[0, a_n-\d_n]$ at its midpoint
$a_{n+1} = (a_n - \d_n)/2$, until it hits either 
$D_0^0$ or $D_0^n$. 
Using $\d_n$ from \eqref{delta}, this is 
\begin{equation}
a_{n+1} = \frac{a_n}{2} - r_n \sqrt{\g-1}
\eqlab{center}
\end{equation}
\begin{figure}[htbp]
\centering
\includegraphics[width=0.5\linewidth]{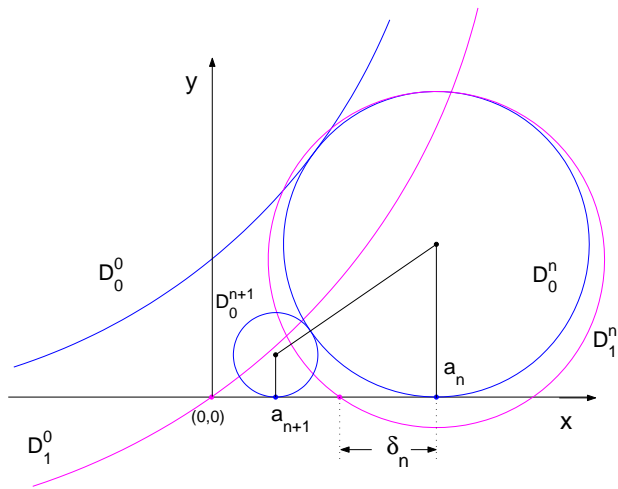} 
\caption{Computing $D^{n+1}_0$ from $D^n_0$.}
\figlab{gap.proof}
\end{figure}

\begin{lemma}
$D^{n+1}_0$ touches $D^n_0$. 
\lemlab{small.touch}
\end{lemma}
\begin{pf}
We determine $r_{n+1}$ from the tangency requirement 
$(a_n-a_{n+1})^2 + (r_n - r_{n+1})^2 = (r_n + r_{n+1})^2$,
or equivalently 
\begin{equation}
r_{n+1} = \frac{(a_n - a_{n+1})^2}{4 r_n},
\eqlab{radius}
\end{equation}
and show that $D^{n+1}_0$ and $D^1_0$ are disjoint:
\[(a_{n+1} - a_0)^2 + (r_0-r_{n+1})^2 > (r_0+r_{n+1})^2\]
Substituting the expression for $r_{n+1}$ from~\eqref{radius} yields
\[a_{n+1} - a_0 > (a_n - a_{n+1})\sqrt{r_0/r_n} > a_n - a_{n+1}\] 
Note that $r_0 > r_n$, since $r_0 \ge r_1$ and 
$r_n$ decreases as $n$ increases. 
Also from \eqref{delta} we have $a_0 = -2 r_0 \sqrt{\g-1}$. 
This together with \eqref{center} renders the inequality 
above true. 
\end{pf}

Our goal is to find the optimal $\g$ for which the algorithm 
terminates in a finite number of steps. This involves solving 
the coupled recurrence relations \eqref{center} and 
\eqref{radius} and imposing the termination condition 
$a_{n+1} - \d_{n+1} \le 0$, which ensures that the 
edge segment is completely covered in iteration step $n+1$. 
Substituting $\d_{n+1}$ from \eqref{delta} yields
\[a_{n+1} - 2 r_{n+1}\sqrt{\g-1} \le 0,\]
which together with \eqref{center} and
\eqref{radius} leads to an system of 
recurrent relations with two variables. 
Next we show how to reduce these recurrence relations
to only one recurrence relation in one variable, 
which is easily solvable.

\subsubsection{Rescaling the gap}
The leftmost segment gap we wish to cover is always 
bounded to the left by $D_0^0$, whose position 
remains unchanged. This suggests a simple way to simplify
the coupled recurrence relations \eqref{center} and
\eqref{radius}: rescale $D_0^n$ at the end of the 
iteration step $n$, so as to ensure $r_n = 1$ at the 
start of the iteration step $n+1$. 
Initially, we scale the disk $D_0^1$ and set
$~r_1^{'} ~=~ r_1 ~/~ r_1 ~=~ 1$ and 
\begin{equation}
a_1^{'} = \frac{a_1}{r_1}
\eqlab{prime.a1} 
\end{equation}
Let $a_n^{'}$ and $r_n^{'}$ denote the scaled variables
at the end of iteration step $n$, with $r_n^{'} = 1$. 
Based on \eqref{center} and \eqref{radius}, we determine 
in iteration step $n+1$ 
\begin{eqnarray}
a_{n+1}^{'} & = & \frac{a_n^{'}}{2} - \sqrt{\g-1} 
\label{scale2} \\
r_{n+1}^{'} & = & \frac{(a_n^{'} - a_{n+1}^{'})^2}{4}
\eqlab{primes}
\end{eqnarray}
Rescale $D_0^{n+1}$ to ensure $r_{n+1}^{'} = 1$. Thus, 
$r_{n+1}^{'} \longleftarrow r_{n+1}^{'} ~/~ r_{n+1}^{'} = 1$
and
\begin{equation}
a_{n+1}^{'} \longleftarrow \frac{a_{n+1}^{'}}{r_{n+1}^{'}}
\eqlab{prime.a}
\end{equation}
Substituting in \eqref{prime.a} the expression for $r_{n+1}^{'}$ from 
\eqref{primes} yields one recurrence relation for $a_n^{'}$ of the form
\begin{equation}
a_{n+1}^{'}  =  F(a_n^{'})
\eqlab{recurrence.prime.a}
\end{equation}
with
\begin{equation}
F(x)  =  4\frac{2 x - 4 \sqrt{\g-1}}{(x + 2 \sqrt{\g-1})^2} 
\eqlab{F}
\end{equation} 
Lemma~\lemref{scaled.nonscaled} establishes the relationship between the 
scaled $a_n^{'}$ and its unscaled correspondent $a_n$:

\begin{lemma}
For each $n$, $~a_n^{'} = a_n ~/~ r_n$ at the end of iteration step $n$.
\lemlab{scaled.nonscaled}
\end{lemma}
\begin{pf}
The proof is by induction on $n$. The base case is $n=1$, which is
clearly true from \eqref{prime.a1}. 
Assume $a_n^{'} = a_n ~/~ r_n$ for any $n \le s$, for some $s > 0$. 
Now we show that $a_{s+1}^{'} = a_{s+1} ~/~ r_{s+1}$.
We use the induction hypothesis $a_s^{'} = a_s/r_s$ in 
\eqref{recurrence.prime.a} to obtain $a_{s+1}^{'} = F(a_s/r_s)$.
From \eqref{center} and \eqref{radius}, we get 
\[ \frac{a_{s+1}}{r_{s+1}} = \frac{4 r_s (a_s/2 - r_s\sqrt{\g-1})}{(a_s - a_{s+1})^2}\]
Substituting again $a_{s+1}$ from \eqref{center} in the expression 
above yields $a_{s+1} / r_{s+1} = F(a_s/r_s) = a_{s+1}^{'}$, 
which proves the lemma.
\end{pf}

\subsubsection{Computing optimal $\g$}
\seclab{optimal.g}
The edge cover algorithm terminates when $D_1^n$ covers all points 
of the uncovered gap, i.e, $a_n^{'} \le \d_n^{'}$.
From \eqref{delta} and the fact that $r_n^{'} = 1$, we derive the
stopping condition
\begin{equation}
a_n^{'} \le 2 \sqrt{\g-1} = \d^{'}
\eqlab{termination}
\end{equation}
Our goal is to determine the optimal $\g$ for which inequality 
\eqref{termination} is satisfied for some finite $n$. 
Clearly, we want 
$a_n^{'}$ to move down to $\d^{'}$, getting closer to 
$\d^{'}$ with each iteration step; that is, 
$a_{n+1}^{'} < a_n^{'}$ for all $n$. 
However, we show that this does not happen for any 
$\g$ and any edge segment $[0, a_1^{'}]$:

\begin{figure}[hptb]
\centering
\includegraphics[width=0.8\linewidth]{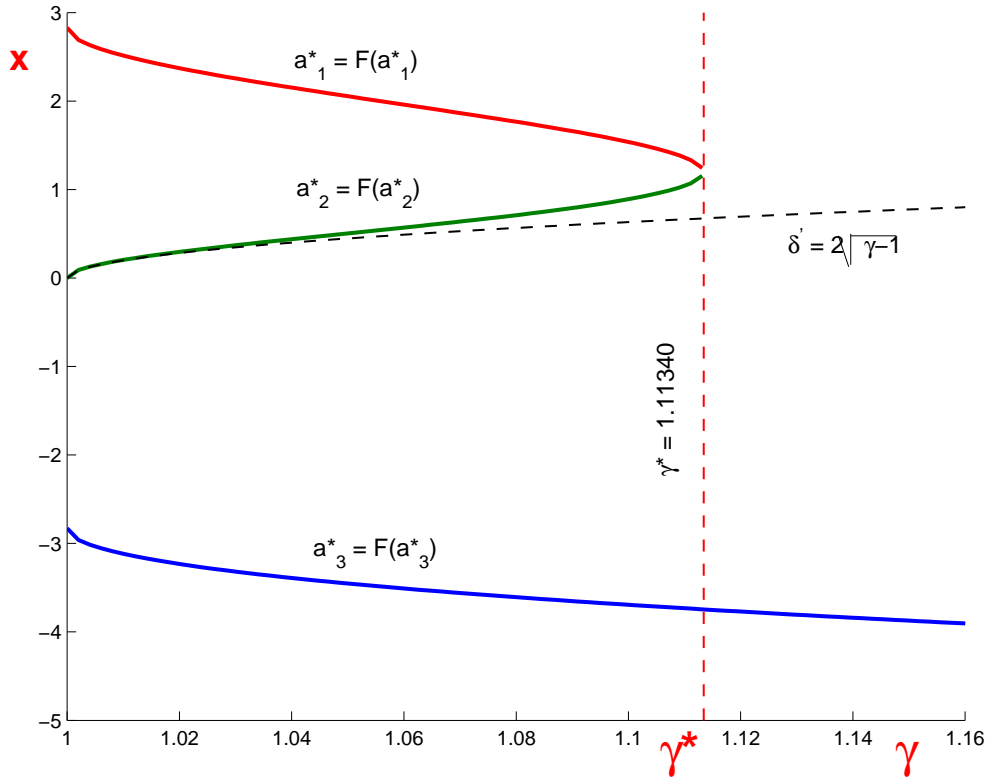} 
\caption{Fixed point.}
\figlab{edge.graph1}
\end{figure}

\begin{theorem}
The algorithm terminates in a finite number of steps only if 
one of the following is true:
\begin{itemize}
\item [(a)] $\g > \g^* = 1.11340$
\item [(b)] $\g < \g^{*}$ and $F(a^{'}_1) < a^{'}_1$ and 
$F^{'}(a^{'}_1) > 1$
\end{itemize}
\theolab{segment.cover}
\end{theorem}
\begin{pf}
The proof consists of three parts. First we show that the 
equation $F(x) = x$ has two positive roots 
$a_1^* > a_2^* > \d^{'}$. Next we 
prove that the iteration procedure $a_{n+1}^{'} = F(a_n^{'})$ 
converges to $a_1^*$, unless one of the two conditions (a) and
(b) stated above is met.
The implication of this is that the edge cover algorithm gets stuck 
at $a_1^*$ and  fails to make any further progress towards $\d^{'}$; 
hence, it never stops.
Finally, we show that under either of the two conditions stated 
in the theorem, the algorithm terminates in a finite number of steps.

Using \eqref{F}, we reduce $x = F(x)$ to a cubic equation 
\begin{equation}
x^3 +  4 \sqrt{\g-1} x^2 + 4 (\g-3) x + 16 \sqrt{\g-1} = 0
\eqlab{fixedpoint}
\end{equation}
which can be solved by use of Cardano's method \cite{g-97}.
Solving for $x$ involves the determinant 
\begin{equation}
\D = -\frac{64}{27}(4 \g^2 - 79 \g + 83)
\eqlab{Delta}
\end{equation}
This quadratic polynomial has one root of interest 
\begin{equation}
\g^* = \frac{79 - 17 \sqrt{17}}{8} = 1.11340
\eqlab{root}
\end{equation}
and a second root outside the domain of interest. 
We omit to show here the complicated expressions for the roots 
of equation \eqref{fixedpoint}.
Figure \figref{edge.graph1} shows with solid curves how 
these roots vary with $\g$. 
The dashed line in Figure \figref{edge.graph1} shows
the finishing point $\d^{'}$. 
Note that for any 
$\g \in [1, \g^*]$, the equation $x = F(x)$ has three real 
roots: two positive roots $a_1^* > a_2^* > \d^{'}$, and one 
negative root. 
Figure~\figref{edge.graph2} shows a magnified view of 
the two positive roots in the vicinity of $\g^* = 1.11340$. 

\begin{figure}[hptb]
\centering
\includegraphics[width=0.76\linewidth]{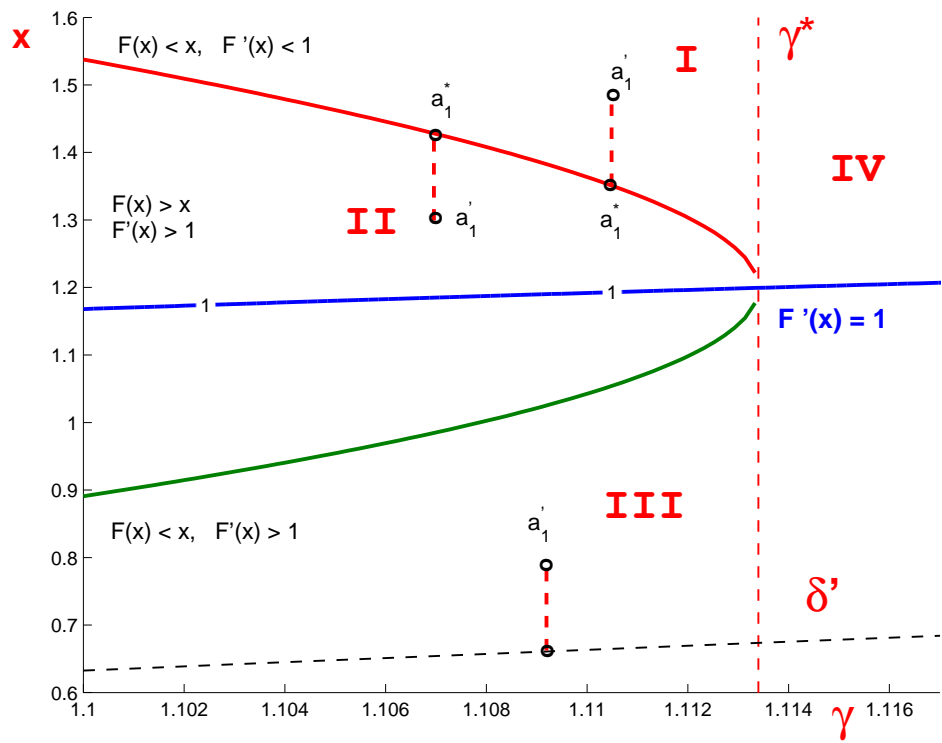} 
\caption{Fixed point.}
\figlab{edge.graph2}
\end{figure}

We now show that for any $\g \in [1, \g^*]$, the iteration
procedure $a_{n+1} = F(a_{n})$ converges to $a_1^*$, unless 
condition (b) of the theorem holds.
From the three regions delimited by the contours of 
the two positive roots $a_1^*$ and $a_2^*$ 
in Figure~\figref{edge.graph2}, observe the following:
\begin{itemize}
\item[(a)] if $F(a_1^{'}) < a_1{'}$, then $a_1^{'}$ lies in
region $I$ above the curve $F(a_1^*) = a_1^*$; therefore, 
$F(a_n^{'}) < a_n^{'}$ for some $N > 0$ and all $n \le N$. 
Also note that $F^{'}(x) < 1$ in a neighborhood containing 
both $a_1^*$ and $a_N^{'}$, which guarantees that $a_n$ converges 
to $a_1^* < a_1^{'}$.
\item[(b)] if $F(a_1^{'}) > a_1{'}$ and 
$F^{'}(a_1^{'}) < 1$, then $a_1^{'}$ lies in region $II$ 
delimited by the contours of the two positive roots; 
therefore, $F(a_n^{'}) > a_n^{'}$ for some $N > 0$ and 
all $ n \le N$.
Again, since $F^{'}(x) < 1$ in a neighborhood containing
both $a_1^*$ and $a_N^{'}$, $a_n$ converges to 
$a_1^* > a_1^{'}$.
\item[(c)] if $F(a_1^{'}) < a_1{'}$ and $F^{'}(a_1^{'}) > 1$, 
then $a_1^{'}$ lies in region $III$ below the curve 
$F(a_2^*) = a_2^*$. Hence, $F(a^{'}_{n}) < a^{'}_{n}$ for 
all $n$ and therefore $a^{'}_n$ reaches $\d^{'}$ in a countable 
number of steps. Also note that the same is true for any
$\g > \g^{*}$ (region $IV$ in Figure~\figref{edge.graph2}).
\end{itemize}
Finally, we show that if the algorithm terminates, then 
$a_n^{'}$ reaches $\d^{'}$ in a finite number of steps. 
In other words, there exists a constant $\e > 0$ such that 
\[a_{n+1} < a_{n} - \e\] 
is satisfied for any iteration step $n$. This is equivalent to 
\begin{equation}
F(x) < x - \e
\eqlab{steps.finite}
\end{equation}
An analysis similar to the 
one of equation~\eqref{fixedpoint} shows that there exists 
$\e > 0$ that satisfies \eqref{steps.finite} for all $x$.
This ensures that the number $n$ of iteration steps is bounded 
above by $(a_1^{'}-\delta^{'})/\e$. 
\end{pf}

Figure~\figref{steps.finite} shows the number of iteration steps 
it takes to cover a segment tangent on its endpoints 
to two unit radius disks tangent to each other.
Note that for any $\g \ge 1.126$, the edge 
segment can be covered in one step only; for any $1.116 \le g < 1.126$, 
the edge segment can be covered in two steps; and so on. As $\g$ approaches 
the critical value $\g^*$, the number of steps increases 
exponentially.

\begin{figure}[hpbt]
\centering
\includegraphics[width=0.7\linewidth]{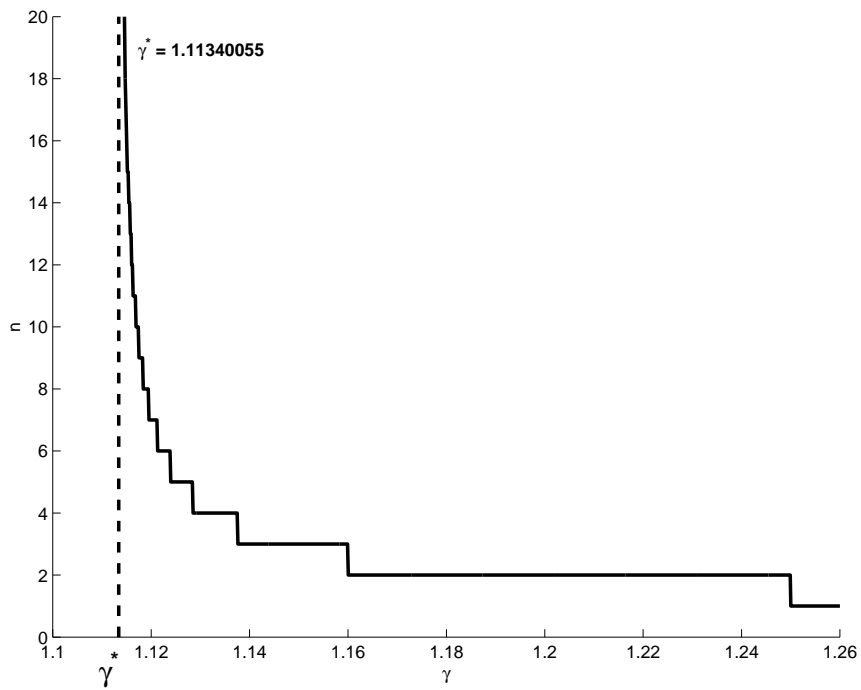} 
\caption{Edge cover ratio $\g$ versus number of iterations.}
\figlab{steps.finite}
\end{figure}

\subsection{Triangular gap partition}

\begin{lemma}
Let $D_0^0$ and $D_0^1$ be two disks tangent to each other
and to an edge segment $S$ at its endpoints. 
If $D_0^2$ covers the intersection point between $D_1^0$ and
$D_1^1$, then a covering produced by the Edge Cover
algorithm for $S$ covers all points of 
the triangular gap delimited by $S$, $D_0^0$ and $D_0^1$.
\lemlab{full.cover} 
\end{lemma}
\begin{pf}
Let $t_n$ be the intersection point between $D^0_1$ and 
$D^n_1$ closer to $S$ in iteration step $n$ 
(see Figure~\figref{gap.apex}). Note that $t_n$ is the apex of 
the triangular gap left uncovered in iteration step $n$, 
which we attempt to cover in iteration step $n+1$. 
If for any $n$, $D^{n+1}_0$ covers $t_n$, then clearly 
the circles $D_1^i$ collectively cover all points of
the original triangular gap. 

\begin{figure}[htbp]
\centering
\includegraphics[width=0.65\linewidth]{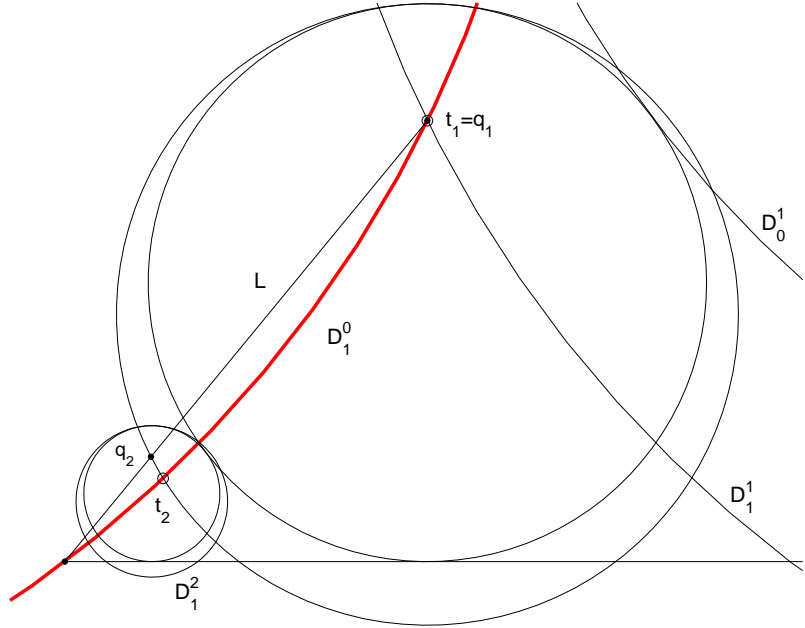} 
\caption{$D^{n}_0$ covers $t_n$ for all $n \ge 1$.}
\figlab{gap.apex}
\end{figure}

As discussed earlier, 
the Edge Cover algoritm terminates only if 
$a_{n+1}^{'} < a_n^{'}$ for all $n$. 
Using Lemma~\lemref{scaled.nonscaled}, 
this is equivalent to 
\begin{equation}
  \frac{a_{n+1}}{a_n} < \frac{r_{n+1}}{r_n}
\eqlab{rate}
\end{equation}
This tells us that $a_n$ decreases at a faster rate than 
$r_n$ with increasing $n$. Let $q_n$ be 
the intersection point between $D_1^n$ and the line $L$ 
that passes through $t_1$ and origin, with 
$q_1 \equiv t_1$. 
Refer to Figure~\figref{gap.apex}. An implication of 
\eqref{rate} is that 
that $q_n$ moves lower inside $D_0^n$ with increasing $n$. 
\notyet
{
In other words, if $y_n$ is the $y$-coordinate of $q_n$, 
then
\[
  \frac{y_{n+1}}{r_{n+1}} < \frac{y_n}{r_n}
\]
}
Therefore, if $q_1$ lies inside $D_0^1$, then $q_n$ lies
inside $D_0^n$ for all $n$.
Also note that $t_n$ always lies below $L$, meaning that  
$D_0^n$ covers $t_n$.
\end{pf}

\begin{lemma}
Let $D_0^0$ and $D_0^1$ be two disks tangent to each other
and to an edge segment $S$ at its endpoints.
If the covering $D_0^i$, $D_1^i$, $i = 0,1,2,\ldots$, 
produced by the edge cover algorithm covers all points of the 
triangular gap $T$ delimited by $D_0^0$, $D_0^1$ and $S$, 
then there exists a partition of $T$ into pieces $T_i$, 
$i = 0,1,2, \ldots,$ such that:
\begin{enumerate}
\item Piece $T_i$ contains $D_0^i$: $T_i \supseteq D_0^i$
\item Piece $T_i$ is contained inside $D_1^i$: $T_i \subseteq D_1^i$
\item The pieces $T_i$ collectively cover $T$: $\cup_{i}T_i \supseteq T$
\end{enumerate}
\end{lemma}
\begin{pf}
Start by assigning points
uniquely covered to the only piece that covers it: 
$T_i = T \cap (D_1^i - \cup_{i \neq j} D_1^j)$. 
Next grow each $T_i$ at a uniform rate from their boundaries, but do 
not permit growth beyond the out-circle boundary. Growth of each set 
is only permitted to consume so-far unassigned points; once a point is 
assigned, it is off-limits for growth. Then 
$T_1, T_2, \ldots$, is a partition of $T$. 
\end{pf}

\section{Pentagon}
A pentagon has $\g_1 = 1/\cos(\pi/5) \approx 1.23607$.
The lower bound provided by Lemma~\lemref{one.angle.lower.bound}
is $\g_\t \approx 1.11803$ (see Table~\tabref{Results}). 
Figure~\figref{partition.square}a shows a partition 
that achieves $\g_\t$, therefore it is optimal.

\begin{figure}[htbp]
\centering
\begin{tabular}{c@{\hspace{0.5in}}c}
\includegraphics[width=0.45\linewidth]{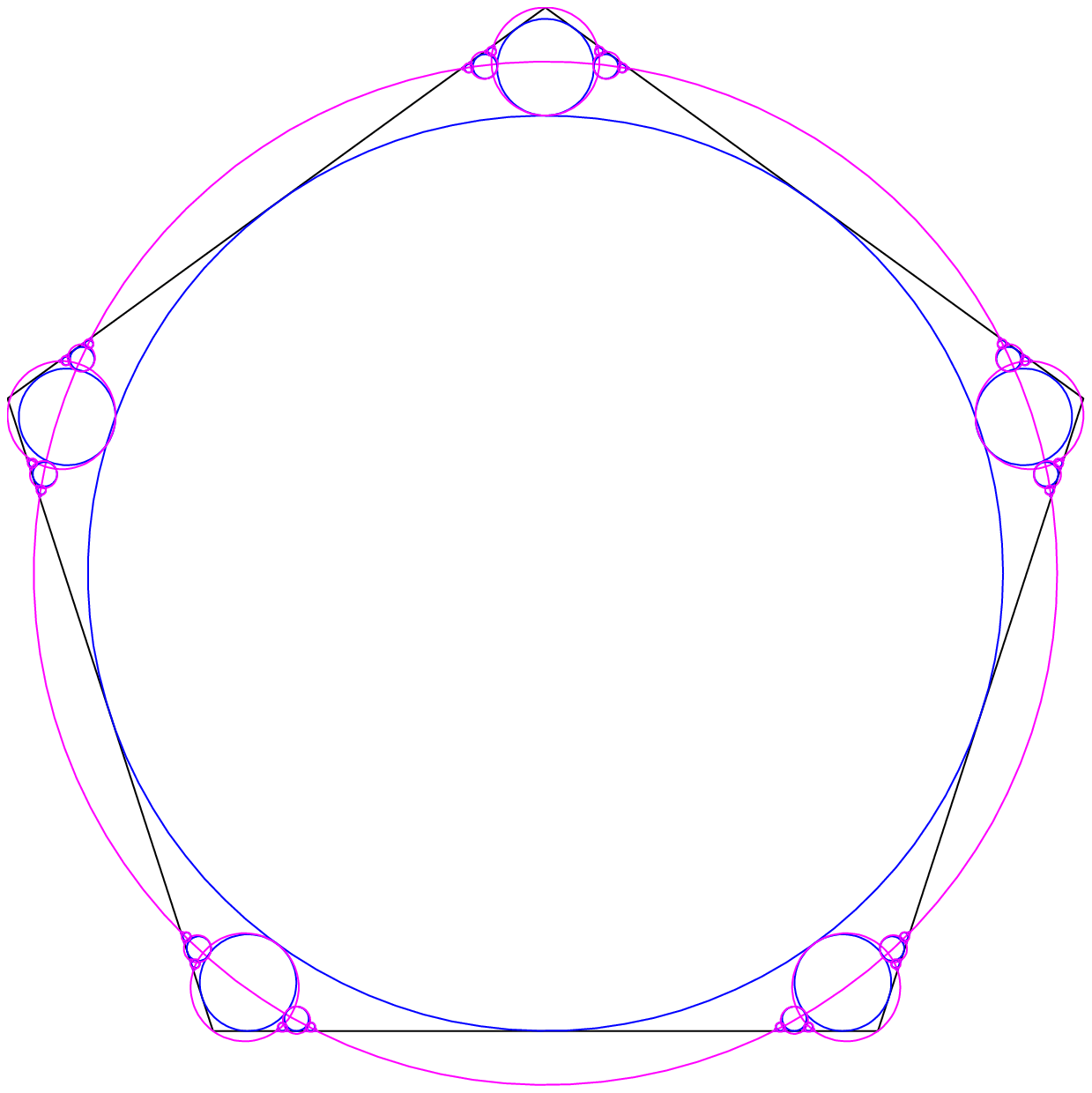} &
\includegraphics[width=0.45\linewidth]{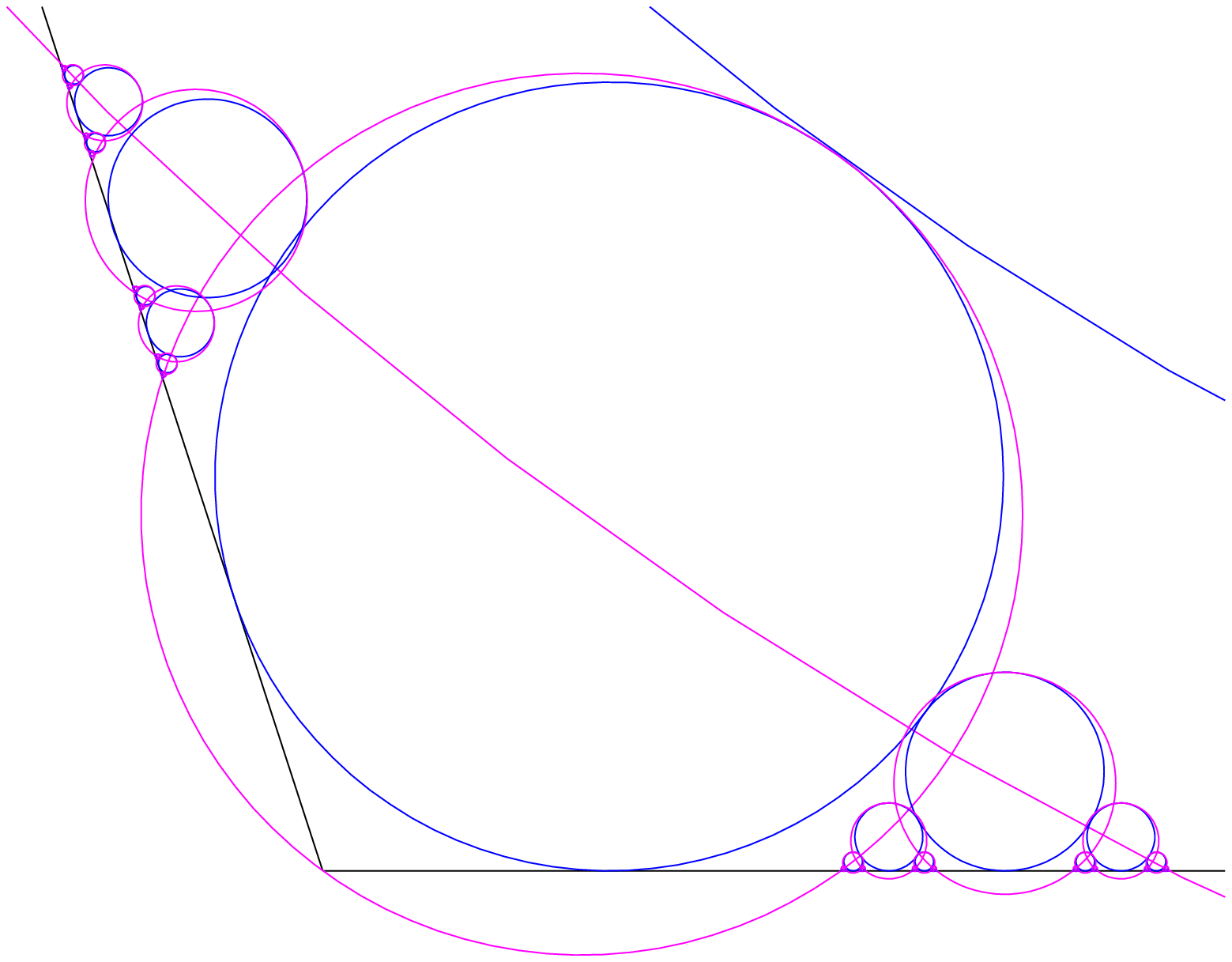}\\
(a) & (b)
\end{tabular}
\caption{(a) Optimal partition of a pentagon (20476 pieces) 
(b) Magnified view of one pentagon corner.}
\figlab{partition.pentagon}
\end{figure}

We start with the pentagon's inscribed circle $D_0^0$ and inflate 
it by $\g_\theta$ to obtain $D_1^0$. In each corner of the pentagon
we nestle five largest possible disks $D_0^1$ and inflate 
each by $\g_\theta$ to obtain $D_1^1$. We choose to make
$D_0^1$ and $D_1^1$ touch each other at the intersection with 
the corner's bisector, so as to create two symmetrical gaps on 
each side of $D_0^1$. 
Cover each of the uncovered edge segments using the edge 
cover algorithm. 
The algorithm uses $12$ iteration steps; therefore, the number 
of partition pieces is $20476 = 5*(2^{12}-1) + 1$, the second term 
counting the big central piece. 
It is easy to verify that $D_0^2$ covers the intersection 
point between $D_1^0$ and $D_1^1$; therefore, conform 
Lemma~\lemref{full.cover}, the algorithm covers all points
interior to the pentagon.

\section{Hexagon and beyond}
A hexagon has $\g_1 = 1/\cos(\pi/6) \approx 1.1547$.
The lower bound provided by Lemma~\lemref{one.angle.lower.bound}
is $\g_\t \approx 1.07735$ (see Table~\tabref{Results}), which is 
below the critical value $\g^*$ of Theorem~\theoref{segment.cover}. 
Intuitively, this means that it is difficult, if not impossible, 
to achieve $\g_\theta$ for $k$-gons for any $k \ge 6$. 
We use the edge cover algorithm described in 
Section~\secref{triangular.gap} to construct partitions 
of $k$-gons, $k \ge 6$, and compute the best $\g$ 
that can be achieved using this algorithm. 

For a fixed $\g$, we partition a $k$-gon into pieces with ratio 
$\g$ as follows. 
As before, we start with the $k$-gon's inscribed disk 
$D_0^0$ and inflate it by $\g$ to obtain $D_1^0$. In each
corner of the $k$-gon we place the largest possible indisk
$D_0^1$ and inflate it by $\g$ to obtain $D_1^1$. We displace
$D_1^1$ along the corner's bisector just enough 
to capture the corner, as shown in 
Figure~\figref{partition.kgon}.
In this way we create two symmetrical triangular gaps on 
each side of $D_0^1$, for a total of $2k$ triangular gaps 
that remain to be covered. Cover each such triangular gap 
using the algorithm from section~\secref{triangular.gap}.

\begin{figure}[htbp]
\centering
\includegraphics[width=0.8\linewidth]{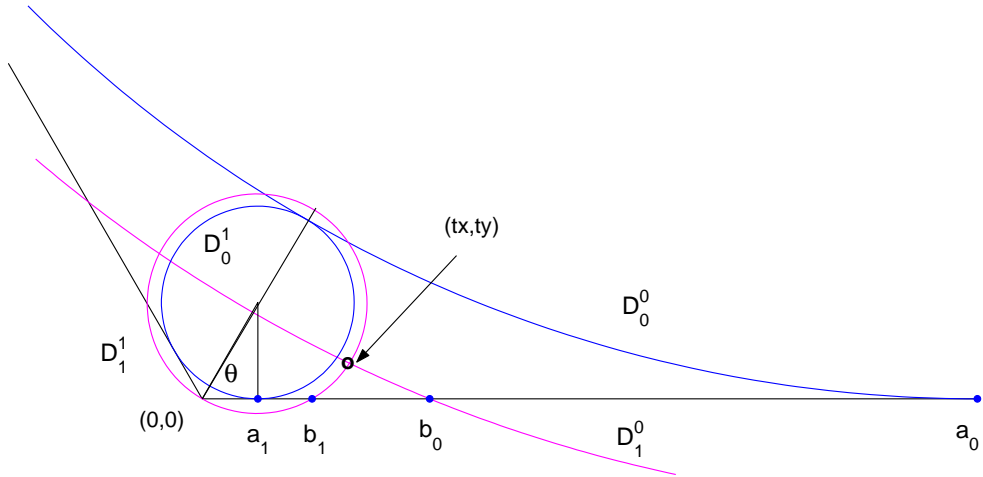} 
\caption{Disk $D_0^0$ nestled in one corner of the $k$-gon.}
\figlab{partition.kgon}
\end{figure}

We now show how to compute the best balancing $\g$ for this 
particular covering. Without loss of generality, we consider
a $k$-gon with unit radius indisk and a coordinate system 
set with the origin at the left corner of the 
bottom horizontal edge.
Let $\t = \pi/2-\pi/k$ denote half of 
the $k$-gon's angle. 
We need to know where $D_1^0$, the inflated 
central circle, cuts the $x$-axis closer to origin:
\begin{equation}
b_0 = \cot(\theta) - \sqrt{\g^2-1}
\eqlab{x2gap}
\end{equation}
Next, we need to compute the corner indisk $D^1_0$:
\begin{equation}
r_1 = \frac{1-\sin(\t)}{1+\sin(\t)}
\eqlab{r1}
\end{equation}
The indisk $D^1_0$ is tangent to the $x$-axis at point 
\begin{equation}
a_1 = r_1 \cot(\t)
\eqlab{a1}
\end{equation}
From this, we can compute the point $b_1$ where $D_1^1$ 
intersects the $x$-axis, closer to origin: 
\[b_1 = 2 r_1 \g \cos(\t)\] 
The edge segment $[b1, b0]$ is covered using the algorithm from 
Section~\secref{triangular.gap}. 
Based on Lemma~\lemref{full.cover}, the gap is fully 
covered if the indisk $D_0^2$ centered at point $(a_2, r_2)$,
with 
\begin{center}
\begin{tabular}{l}
$a_2 = (b_0 + b_1)/2$ \\
$r_2 = (a_1 - a_2)^2/4 r_1$,
\end{tabular}
\end{center}
covers the apex of the triangular gap. 
Note that the initial 
scaled gap value used in equation~\eqref{recurrence.prime.a} 
is $a_0^{'} = (b_0 - a_1)/r_1$. Substituting the expressions
for $b_0$ \eqref{x2gap}, $r_1$ \eqref{r1} and $a_1$ \eqref{a1}, 
this expands to
\[a_0^{'} = \frac{1+\sin(\t)}{1-\sin(\t)}
            (\cot(\theta) - \sqrt{\g^2-1}) - \cot(\t)\]
We now solve for $\g$ that satisfies 
\begin{equation}
\left\{
\begin{array}{l}
F(a_0^{'}) < a_0^{'} \\
F^{'}(a_0^{'}) > 0 \\
(a_2-t_x)^2 + (r_2 - t_y)^2 \le r_2^2 
\end{array}
\right.
\eqlab{kopt}
\end{equation}
where $(tx, ty)$ is the intersection point 
between outcircles $D_1^0$ and $D_1^1$.
Conform Theorem~\theoref{segment.cover}, the first two
inequalities in \eqref{kopt} ensure that the edge cover algorithm 
terminates in a finite number of steps. 
Conform Lemma~\lemref{full.cover}, the third inequality in 
\eqref{kopt} ensures that the algorithm covers the entire triangular 
gap. 

\begin{figure}[htbp]
\centering
\includegraphics[width=0.6\linewidth]{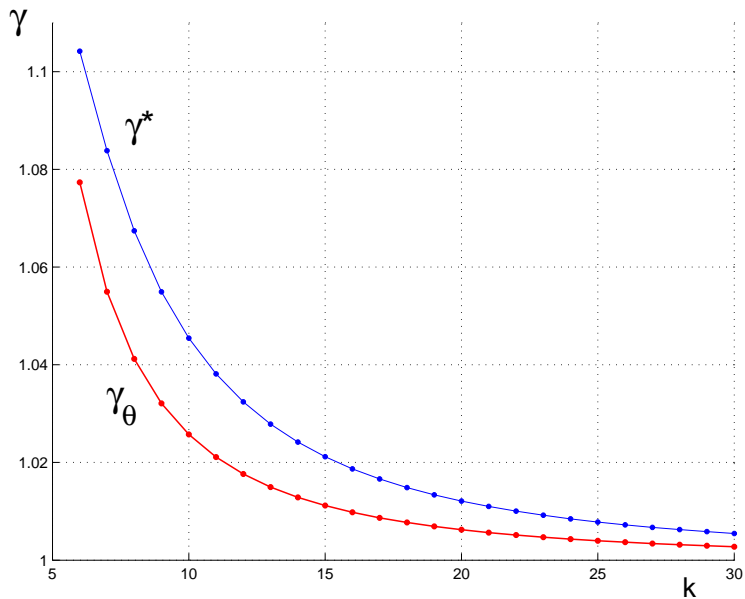} 
\caption{(a) Optimal ratio $\g^*$ and single-angle lower bound $\g_\theta$
versus number of vertices $k$ of regular $k$-gons.}
\figlab{partition.kopt}
\end{figure}

Solving \eqref{kopt} for $\g$ and $k = 6, 7$ and $8$ 
yields the ratio values shown in Table~\tabref{Results}. 
The top curve in Figure~\figref{partition.kopt} 
shows how the ratio $\g^{*}$ that satisfies~\eqref{kopt} varies with $k$. 
The bottom curve represents the single-angle lower bound
ratio $\g_\theta$, which is best any algorithm could achieve. As is 
clear from Figure~\figref{partition.kopt}, the ratio achieved by our 
algorithm is close to the optimal. 

\section{Discussion}
We leave open the question of whether optimal paritions can be
achieved for $k \ge 6$ with a finite number of pieces.

\end{document}

%% file: mac.tex
\newenvironment{pf}{\unskip{\bf Proof:}}{\unskip{\hfill $\Box$}}

\newcommand{\lemlab}[1]{\label{lemma:#1}}
\newcommand{\theolab}[1]{\label{theo:#1}}

\newcommand{\eqlab}[1]{\label{eq:#1}}

\newcommand{\tablab}[1]{\label{tab:#1}}
\newcommand{\figlab}[1]{\label{fig:#1}}
\newcommand{\seclab}[1]{\label{section:#1}}

\newcommand{\lemref}[1]{\ref{lemma:#1}}

\newcommand{\theoref}[1]{\ref{theo:#1}}

\newcommand{\tabref}[1]{\ref{tab:#1}}
\newcommand{\figref}[1]{\ref{fig:#1}}
\newcommand{\eqref}[1]{(\ref{eq:#1})}
\newcommand{\secref}[1]{\ref{section:#1}}

\newtheorem{theorem}{Theorem} 
\newtheorem{lemma}[theorem]{Lemma}

{\catcode`\@=11
\gdef\setft#1#2#3{%
\def\@oddfoot{
{\setbox0=\hbox{#1}
\setbox1=\hbox{#3}
\ifdim\wd0>\wd1
\dimen0=\wd0
\box0\hfil#2\hfil\hbox to\dimen0{\hfil\hfil\box1}
\else \dimen0=\wd1
\hbox to\dimen0{\box0\hfil }\hfil#2\hfil\box1 \fi
}}} }

\def\complaint#1{}
\def\withcomplaints{
\newcounter{mycomplaints}
\def\complaint##1{\refstepcounter{mycomplaints}%
\ifhmode%
\unskip%
{\dimen1=\baselineskip \divide\dimen1 by 2 %
\raise\dimen1\llap{\tiny -\themycomplaints-}}\fi%
\marginpar{\tiny [\themycomplaints]: ##1}}%
}

%% file: Nonconvex.bbl
\begin{thebibliography}{Gul97}

\bibitem[DO03]{mo-cp-03}
M.~Damian and J.~O'Rourke.
\newblock Partitioning regular polygons into circular pieces \protect{I}:
  Convex partitions.
\newblock {\em Proc. 15th Canad. Conf. Comput. Geom.}, pages 43--46.
\newblock 2003.
\newblock 
  \protect\url{http://arXiv.org/abs/cs.CG/0304023}.


\bibitem[Gul97]{g-97}
J.~Gullberg.
\newblock {\em Mathematics from the birth of numbers}.
\newblock W.W.~Norton, New York, 1997.

\end{thebibliography}
